\documentclass[pre,superscriptaddress,
showpacs, twocolumn
]{revtex4}
\usepackage{graphicx,amsmath,amssymb,natbib}
\begin{document}
\title{Nematic Droplets in Aqueous Dispersions of Carbon Nanotubes}

\author{Nicolas Puech}
\author{Eric Grelet}
\author{Philippe Poulin}
\affiliation{Centre de Recherche Paul-Pascal, CNRS-Universit\'{e}
Bordeaux 1, 115 Avenue Schweitzer, 33600 Pessac, France}
\author{Christophe Blanc}
\affiliation{Laboratoire des Collo\"ides, Verres et Nanomat\'eriaux,
CNRS-Universit\'{e} Montpellier II, Place E. Bataillon, 34090
Montpellier, France}
\author{Paul van der Schoot}
\affiliation{Faculteit Technische Natuurkunde, Technische
Universiteit Eindhoven, Postbus 513, 5600 MB Eindhoven, Netherlands,
and Instituut voor Theoretische Fysica, Universiteit Utrecht,
Leuvenlaan 4, 3584 CE Utrecht, Netherlands}

\date{\today}

\begin{abstract}
Aqueous dispersions of exfoliated, bile-salt stabilized single-wall
carbon nanotubes exhibit a first order transition to a nematic
liquid-crystalline phase. The nematic phase presents itself in the
form of micron-sized nematic droplets also known as tactoids, freely
floating in the isotropic host dispersion. The nematic droplets are
spindle-shaped and have an aspect ratio of about four, irrespective
of their size. We attribute this to a director field that is uniform
rather than bipolar, which is confirmed by polarization microscopy.
It follows that the ratio of the anchoring strength and the surface
tension must be about four, which is quite larger than predicted
theoretically but in line with earlier observations of bipolar
tactoids. From the scatter in the data we deduce that the surface
tension of the coexisting isotropic and nematic phases must be
extremely low, that is, of the order of nN/m.
\end{abstract}


 \maketitle

Carbon nanotubes or CNTs are colloidal particles with a very large
aspect ratio, typically in the range from many tens to hundreds up
to even thousands. Hence, it is not surprising that, provided they
are properly stabilized against aggregation, fluid dispersions
containing CNTs exhibit an Onsager-type isotropic-nematic transition
\cite{CNT1,CNT2,CNT3,CNT4,CNT5,CNT6}. This happens at concentrations
in excess of a critical value that depends on the aspect ratio of
the rods. The relevant concentration scale here is the volume or
packing fraction because the driving force for the spontaneous
alignment is the anisotropic volume exclusion between the particles.
For the nematic to become stable the volume fraction of CNTs should
be in excess of a few times the reciprocal of some average of their
aspect ratios \cite{Onsager49}. It follows that the nematic
transition must occur at very low concentrations of, say, one per
cent of CNTs. This, by and large, is in agreement with experimental
observation, allowing for instance for the effects of polydispersity
\cite{CNT1,CNT2,CNT3,CNT4,CNT5,CNT6}.

\begin{figure}
\includegraphics[width=0.45\textwidth]{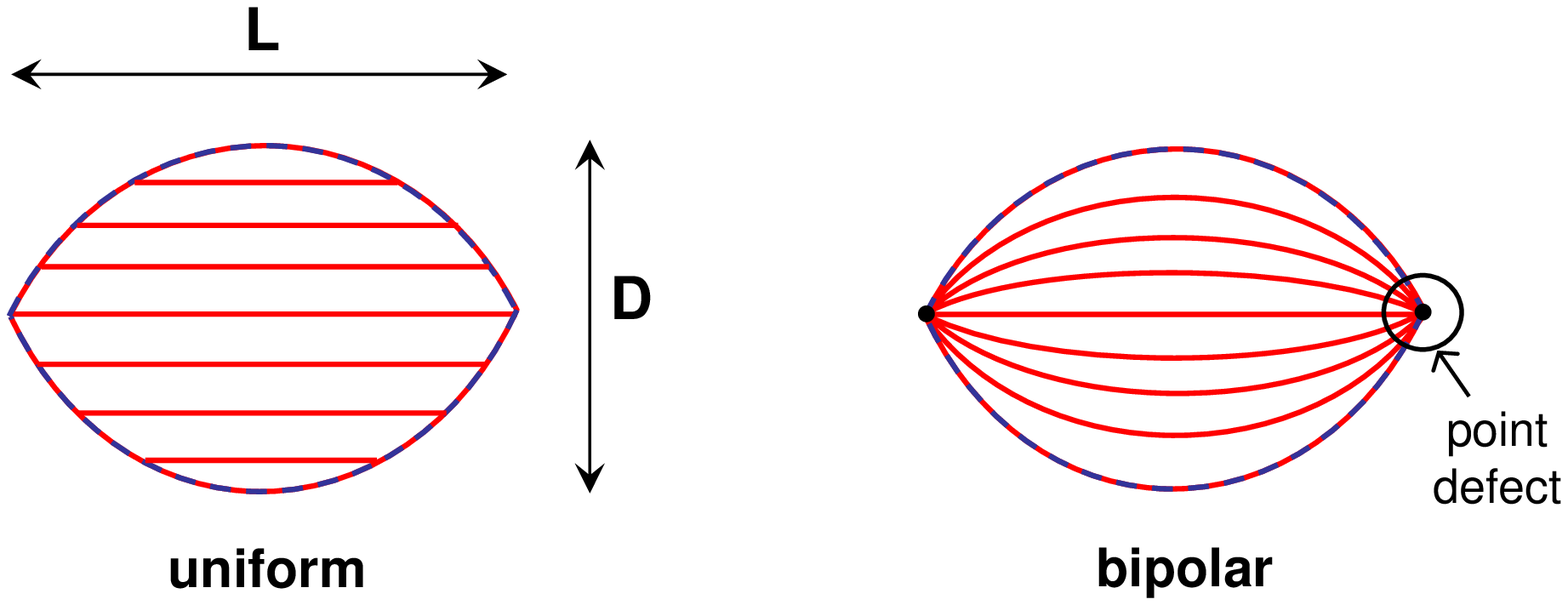}
\caption{\label{Scheme} (Color online) Schematic representation of
 the shape and director field in a uniform
 (left) and a bipolar (right) tactoid. The shape of the uniform tactoid
 is determined by the anchoring strength; the larger it is relative to the
 interfacial tension the more elongated the tactoid becomes. The shape of
 a bipolar tactoid is determined by the elastic deformation favoring
 an elongated shape, an effect making the director field more
 uniform, and by the surface tension that favors as small a surface
 area as possible. The crossover between the two types of director
 field occurs when anchoring energy of the one and elastic energy of
 the other are equal \cite{Prinsen03}.}
\end{figure}

Often, before isotropic-nematic phase separation occurs on a
macroscopic scale in dispersions of elongated colloidal particles,
the nematic phase establishes itself in the form of droplets called
tactoids. Tactoids have been observed in many dispersions, such as
tobacco mosaic virus \cite{Bernal41,Maeda97}, boehmite rods
\cite{vanBruggen98}, poly(butyl glutamate), self-assembled
chromonics \cite{Lavrentovich05}, \textit{fd} virus \cite{Fraden01},
f-actin \cite{Tang07}, and vanadium pentoxide
\cite{Kaznacheev02,Sonin98}. These droplets have in common their
unusual elongated, spindle-like shape. This shape can be explained
by the preferential planar anchoring of the nematic director at the
interface with the isotropic phase. The competition between surface
tension and the elastic deformation of the bipolar director field
that accommodates this preferential planar anchoring plausibly
determines the optimal aspect ratio (Figure \ref{Scheme}). If the
director field of a tactoid is indeed bipolar with field lines
connecting two boojum surface defects, then its shape, as described,
e.g., by the aspect ratio or the tip angle, depends on its physical
dimensions because the surface and bulk elastic energies scale
differently with droplet size. This seems to be the case in all
systems investigated so far. Indeed, information on the ratio of (an
average of) the elastic constants and the surface tension can be
obtained from the measured relation between, say, the aspect ratio
and length of the droplets
\cite{Kaznacheev02,Kaznacheev03,Prinsen03,Prinsen04,Tang07}.

In this letter, we show that aqueous dispersions of
surfactant-stabilized carbon nanotubes deviate from the usual
picture of a bipolar director field. The tactoids that we observe in
these dispersions are also quite elongated, but display a uniform
director field as evidenced by polarization microscopy. This ties in
with our finding that the aspect ratio of  the tactoids is
independent of their size, at least for the one decade range of
sizes present in our samples. Theoretically, the aspect ratio of a
tactoid is independent of its size if it is dictated by the surface
tension anisotropy, i.e., the ratio of the anchoring strength and
the surface tension \cite{Prinsen03}. This is plausible if the
director field is uniform and as a result of that the anchoring
conditions at the surface of the drops are suboptimal. It is because
the interfacial and anchoring free energies both scale with the area
of a drop that its shape can only  be a function of the ratio of the
anchoring strength and the surface tension. From our observations we
find that the anchoring strength is about four times larger than the
bare surface tension. We put forward that the surface tension must
be very small indeed, possibly as low as 1 nN/m.

We prepared our CNT tactoid dispersions from an aqueous suspension
of single-wall carbon nanotubes (furnished by Elicarb batch K3778)
and dispersed by bile salts, at the respective concentrations of
0.5\% w/w CNTs and 0.5\% w/w bile salts. To exfoliate the CNT
bundles, sonication was applied to the suspension for a period of
three hours. A purification process by selective centrifugation was
then performed on the CNT suspensions. After removing CNT aggregates
by centrifugation at low speed (30 min, 3500 rpm), the longest
carbon nanotubes exhibiting some entanglements and defects were
removed by ultracentrifugation (45 min, 45000 rpm). A second
ultracentrifugation (180 min, 45000 rpm) was applied to the CNT
suspensions to obtain the nematic liquid crystalline phase, which
appeared as a black pellet on the bottom of the centrifugation tube.
Finally, CNT tactoids were obtained by diluting the nematic phase up
to the coexistence region with the isotropic phase. Samples of a few
$\mu$m thick were prepared at the isotropic-nematic phase
coexistence between cover slip and glass slide. The CNT tactoids
were observed by optical microscopy at different magnifications
between crossed polarizers.

\begin{figure}
\includegraphics[width=0.45\textwidth]{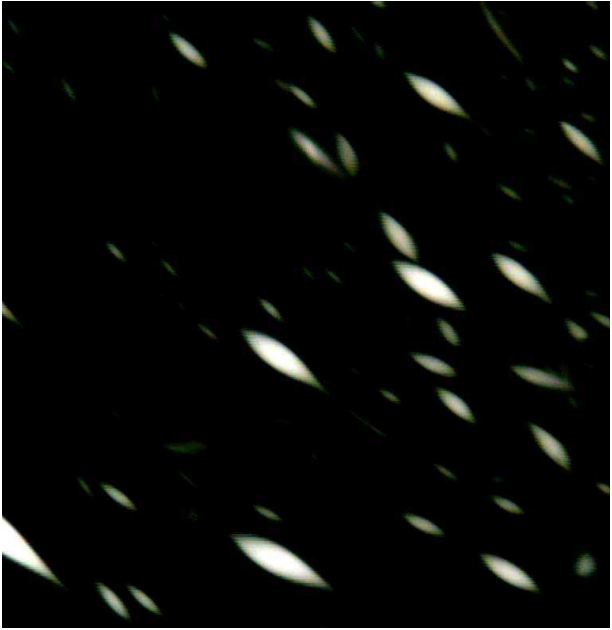}
\caption{\label{ZooTactoids} Polarization microscopic image with
crossed polarizers, showing a collection of tactoids in aqueous
dispersion of bile-salt stabilized single-wall carbon nanotubes. The
image size is 97~$\mu$m~$\times$~100~$\mu$m. }
\end{figure}

\begin{figure}
\includegraphics[width=0.45\textwidth]{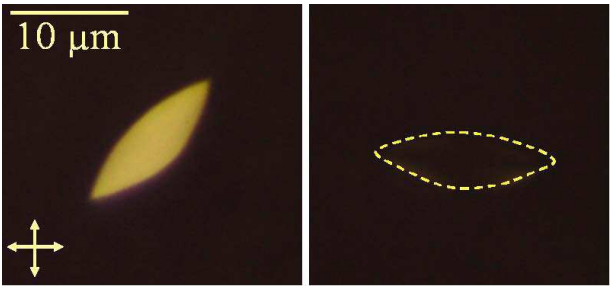}
\caption{\label{TactoidP&A} (Color online) Observation by
polarization microscopy of a tactoid with the main body axis at 45
degrees relative to the polarizers (left) and along one
 of the polarizers (right). Indicated by the dashed line
 is the outline of the tactoid if oriented at 45 degrees 
 with respect to the polarizers. 
  }
\end{figure}

Figure \ref{ZooTactoids} shows an image by polarization microscopy
of the kind of tactoids that we find in our carbon nanotube
dispersions. The background is dark because the tactoids float in
the coexisting isotropic liquid phase. The tactoids are indeed quite
elongated and have the typical spindle-like shape. A perfect
extinction is observed  when the tactoids are aligned along the
polarizer or analyzer direction, demonstrating that the director
field is uniform, as shown for a typical tactoid in figure
\ref{TactoidP&A}. A further study under confocal Raman
micro-spectroscopy also confirms that the director filed is aligned
along the tactoid long axis \cite{Puech}, which coincides with the
main optical axis.

\begin{figure}
\includegraphics[width=0.48\textwidth]{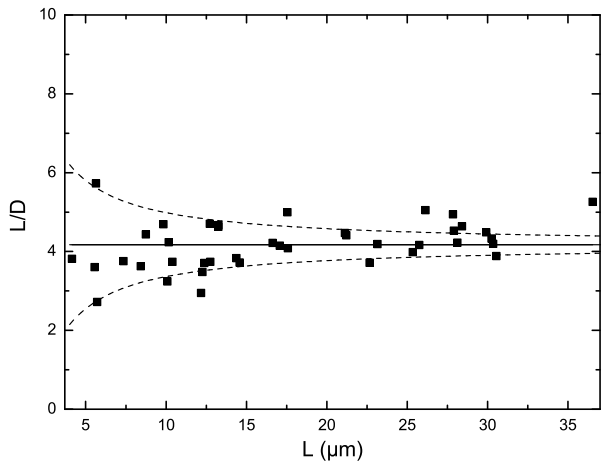}
\caption{\label{PlotLD} Aspect ratio of the tactoids versus their
length in microns. Full squares: experimental data points. Drawn
 line: average value deduced from the experiments. Dashed
lines: predicted standard deviation presuming that the surface
tension is equal to 0.5 nN/m. }
\end{figure}

\begin{figure}
\includegraphics[width=0.45\textwidth]{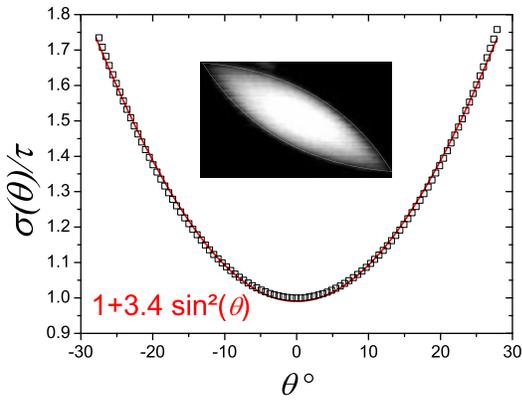}
\caption{\label{wulff} (Color online) Inverse Wulff construction
applied to a tactoid providing the form of surface tension
anisotropy. The experimental data points (indicated by the squares)
are well fitted with a Rapini-Papoular anchoring expression (drawn
line).}
\end{figure}

In figure \ref{PlotLD} the aspect ratio of a large collection of
tactoids is plotted against their length, which, within experimental
error, is constant as advertised. Apparently, the mean aspect ratio
of the tactoids is about four for tactoids up to 36 $\mu$m in
length. This result is consistent with a uniform director field in
the droplets, as the aspect ratio of the drops is then dictated by
the ratio of the anchoring strength and the surface tension
\cite{Prinsen03}. By applying an inverse Wulff construction
\cite{Wulff01} to the shape of a typical tactoid, we have been able
to probe the anisotropy of the surface free energy, $\sigma$. Shown
in figure \ref{wulff} is $\sigma(\theta)/\tau$, where $\tau$ is the
bare surface tension and $\theta$ the angle between the surface
tangent and the director field. We find that this angle dependence
is consistent with the often-used Rapini-Papoular model for the
surface free energy, i.e., $\sigma = \tau ( 1+ \omega \sin^2
\theta)$ with $\omega$ the ratio of the anchoring energy and the
surface tension \cite{RapiniPapoular69}. For the tactoid shown in
the figure, we obtain a value for the dimensionless anchoring
strength of $\omega \approx 3.4$. From the Wulff construction, we
can in fact predict the aspect ratio of a nematic drop that within a
Rapini-Papoular model must be equal to $2\sqrt{\omega}$ if $\omega
\geq 1$ \cite{Prinsen03}. It follows from figure \ref{PlotLD} that
for our coexisting isotropic and nematic phases of CNTs in water,
$\omega = 4\pm1$. This is quite larger than theoretical predictions
according to which $0.5 \leq \omega \leq 1.5$, at least for
cylindrical particles interacting via a harshly repulsive potential
\cite{Chen92,Chen95,Harlen99,vanderSchoot99}.

Additional information that we can deduce from figure \ref{PlotLD}
can be obtained by realizing that according to recent calculations
\cite{Prinsen03}, a tactoid changes its director field from a
uniform to a bipolar one if its volume
$V\approx(4\pi/3)(L/2)(D/2)^2$ is larger than
$(K/\tau)^3(6.25/\omega)^{18/5}$, with $K$ an average of the Frank
elastic constants. It has to be noted that this crossover has so far
only been observed in computer simulations \cite{Trukhina09}. It
seems that even the largest tactoid in our samples with a length of
36 $\mu$m, has a uniform director field. This implies that a lower
bound for the ratio $K/\tau$ of the nematic of CNTs must be
approximately 5 $\mu$m. Interestingly, this lower bound is
comparable to values found for vanadium pentoxide and {\it{fd}}
virus \cite{Kaznacheev02,Prinsen03,Prinsen04}, although that
tactoids in these two systems do exhibit a bipolar field if larger
than a few $\mu$m. It is not quite clear why single-wall CNTs behave
so differently from other types of elongated particle. Indeed, CNTs
exhibit a phase behavior that is in good agreement with the behavior
expected for bulk suspensions of rod-like particles \cite{CNT3}.
Nevertheless the structure of tactoids does result from a delicate
interplay of bulk elastic and surface properties of the coexisting
isotropic and nematic phases. These properties are known to quite
sensitively depend on molecular details such as a the degree of
bending flexibility and the type and the strength of interactions
involved in stabilizing them in suspension. 
Another issue is also the influence of polydispersity, which is
known to be large. CNTs are polydisperse and a small fraction of
very long or very short particles could affect the surface
properties with a little effect on the bulk behavior. Of course,
this is speculative and further work will be needed to confirm
whether or not the size distribution of the CNTs confined at the
isotropic-nematic interface differs from that in bulk.

Finally, the scatter in the data of figure \ref{PlotLD} potentially
provides physical information because in part it must be caused by
thermal fluctuations of the aspect ratio of the tactoids. The Wulff
construction provides only the optimal aspect ratio but does not
give an indication of its variance. Using a simple scaling Ansatz
based on the Rapini-Papoular surface free energy \cite{Prinsen03}
and making use of the equipartition of free energy, we find that the
standard deviation of the aspect ratio must approximately be equal
to $(k_B T/\tau L^2)^{1/2}\omega^{3/4}$, so depend on the size of
the tactoids. In figure \ref{PlotLD} we have indicated around the
estimated average aspect ratio the standard deviation presuming a
surface tension of 0.5 nN/m. The prediction follows the magnitude
and size dependence of the experimental scatter in the data
reasonably well. Obviously, we should not over interpret this result
because we have ignored any influence of the intrinsic error in the
size measurement of the tactoids.

If we accept the very small value of the surface tension that we
find at face value, then it is very much smaller than values in the
range of tenths to tens of $\mu$N/m typically found for coexisting
isotropic and nematic phases in dispersions of rod-like particles
\cite{Gray02}. Values of nN/m have been found for coexisting
isotropic and nematic phases but only in dispersions of colloidal
platelets \cite{Vanderbeek06}. Clearly, the scaling theory does not
give the numerical pre-factor, and this could increase the found
surface tension by another factor of, say, ten. Still, this by no
means takes our value within the range of the other experimentally
found values. One might of course conclude from this, that the
presumption that the scatter in the data is dominated by thermal
fluctuations must be wrong. Indeed, some flow can occur just after
sample preparation, which can induce droplet alignment and some
shape distortion. The latter is clearly seen in the largest CNT
tactoids, see also figure \ref{ZooTactoids}. Explicit surface
tension measurements, e.g., by the capillary rise method
\cite{Vanderbeek06} on macroscopic interfaces between coexisting
isotropic and nematic phases are necessary to confirm our finding,
but these are outside the scope of the present paper.

In conclusion, we find nematic tactoids of aqueous dispersions of
surfactant-stabilized single-wall carbon nanotubes that have a
uniform director field. Our observation accounts that the aspect
ratio of about four is independent of the size of these tactoids at
least for lengths up to 36 $\mu$m. The crossover to a bipolar
director field must occur for sizes much larger than this value, but
we have not been able to confirm this.

\begin{acknowledgments}
PvdS gratefully acknowledges the hospitality and a supporting grant from Universit\'{e} Bordeaux 1.
\end{acknowledgments}

\newpage

\end{document}